# TEACHING SPREADSHEET DEVELOPMENT USING PEER AUDIT AND SELF-AUDIT METHODS FOR REDUCING ERRORS

*Chadwick D., Sue R.,*

*Information Integrity Research Centre,*

*School of Computing & Mathematical Sciences,*

*University of Greenwich, London SE10 9LS, United Kingdom*

Phone:+44 0208 8331 8509    Fax: 0208 8331 8665    Email:cd02@gre.ac.uk

**Abstract**

*Recent research has highlighted the high incidence of errors in spreadsheet models used in industry. In an attempt to reduce the incidence of such errors, a teaching approach has been devised which aids students to reduce their likelihood of making common errors during development. The approach comprises of spreadsheet checking methods based on the commonly accepted educational paradigms of peer assessment and self-assessment. However, these paradigms are here based upon practical techniques commonly used by the internal audit function such as peer audit and control and risk self-assessment. The result of this symbiosis between educational assessment and professional audit is a method that educates students in a set of structured, transferable skills for spreadsheet error-checking which are useful for increasing error-awareness in the classroom and for reducing business risk in the workplace.*

## 1.0 THE PROBLEM OF SPREADSHEET ERRORS

The world-wide problem of spreadsheet errors and the need to act to prevent, detect and minimise the effect of errors has been widely reported both in industry studies and in more formal academic research studies.

### 1.1 The Industry Experience

The Business Modelling Group at KPMG Consulting have been only too aware of the problems of spreadsheet errors and frequently cite the findings of a KPMG survey of financial models based on spreadsheets. The survey findings, reported in an article in Internal Auditing by Chadwick, D. [1], found that:

- 95% of models were found to contain major errors (errors that could affect decisions based on the results of the model),
- 59% of models were judged to have 'poor' model design,
- 92% of those that dealt with tax issues had significant tax errors,
- 75% had significant accounting errors.

These findings have been reinforced by a statement in the same article by David Finch, Head of Internal Audit at Superdrug plc, who believed that people shouldn't be too surprised at the high rate of spreadsheet errors as:

*"The use of spreadsheets in business is a little like Christmas for children. They are too excited to get on with the game to read or think about the 'rules' which are generally boring…. There is often little control over end user developments in spreadsheets with little if any standardisation in development processes by users in different departments, little risk analysis and a general assumption that models, on which important business decisions are made, are accurate"*

The UK Customs and Excise department collects VAT on behalf of the British government. Raymond Butler, a computer auditor in the department, has written extensively on the frequency of errors found in spreadsheets used by businesses to calculate the amount of VAT that they should pay.

**Butler and his team have done much work on risk assessment methods to enable them to determine which VAT spreadsheets have a likelihood of error and need to be audited** ; Butler R. [13]. One of the questions the Customs & Excise audit team ask themselves is whether the organisation for whom the development is being made have an adequate policy governing development, testing and use of spreadsheet models and applications. They also

**investigate**

whether there is evidence that this policy is observed and enforced.

Butler states that there are many examples of errors found in both field audits and experiments which show that good development practice is rarely codified into business procedures and even when it is, the rules and restrictions it requires are not followed to any significant degree.

### 1.2 The Academic Research Experience

For the past 15 years, Professor Raymond Panko at the University of Hawaii has been studying and writing prolifically upon the phenomenon of spreadsheet error. In addition to compiling bodies of data from research establishments and businesses around the world he has conducted his own experiments with students. It is not possible to report upon all of Panko's findings in this paper but a recent paper of his shows a comparison of findings from several educational studies (see fig.1.1).

Fig.1.1 Spreadsheet Development Experiments : Panko R. [12]

| Study | Year | Sample | Subjects | Spread sheets | % with Errors | Cell Error Rate (CER) |
|---|---|---|---|---|---|---|
| Brown & Gould | 1987 | ED | 9 | 27 | 63% | NR |
| Olson & Nilsen (1,2) | 1987-1988 | ED | 14 | 14 | NA | 21% |
| Lerch (1,2) | 1988 | ED | 21 | 21 | NA | 9.3% |
| Hassinen (2) on paper | 1988 | Ugrad | 92 | 355 | 55% | 4.3% |
| Hassinen (2) online | 1988 | Ugrad | 10 | 48 | 48% | NR |
| Janvrin & Morrison (3) Study 1, alone | 1996 | Ugrad | 78 | 61 | NR | 7% to 10% |
| Janvrin & Morrison (3) Study 1, dyads | 1996 | Ugrad | 88 | 44 | NR | 8% |
| Janvrin & Morrison (3) Study 2, alone | 1996 | Ugrad | 88 | 88 | NR | 8% to 17% |
| Kreie (post test) | 1997 | | 73 | 73 | 42% | 2.5% |
| Teo & Tan (4) | 1997 | Ugrad | 168 | 168 | 42% | 2.1% |
| Panko & Halverson, alone | 1997 | Ugrad | 42 | 42 | 79% | 5.6% |
| Panko & Halverson, dyads | 1997 | Ugrad | 46 | 23 | 78% | 3.8% |
| Panko & Halverson, tetrads | 1997 | Ugrad | 44 | 11 | 64% | 1.9% |
| Panko & Sprague (4) | 1999 | Ugrad | 102 | 102 | 35% | 2.2% |
| Panko & Sprague (4,5) | 1999 | MBA (NE) | 26 | 26 | 35% | 2.1% |
| Panko & Sprague (4,6) | 1999 | MBA ED) | 17 | 17 | 24% | 1.1% |
| Panko & Halverson, monads | 2000 | Ugrad | 35 | 35 | 86% | 4.6% |
| Panko & Halverson, triads | 2000 | Ugrad | 45 | 15 | 27% | 1.0% |
| Total Sample | | | 998 | 1170 | 51% (7) | |

NR = not reported  ED = experienced developer
NE = not very experienced with development at work  Ugrad = undergraduate students

(1) Measured errors before subject had a chance to correct them: (2) Only measured error rate in formula cells : (3) Only measured error rate in cells linking spreadsheets : (4) Wall Task designed to be relatively simple and free of domain knowledge requirements : (5) MBA students with little or no development experience (6) MBA students with considerable development experience : (7) Weighted average

Panko's figures show that, over the given experiments, 51% of the spreadsheets contained errors. Of course no mention is made of the type of errors and their severity in terms of spreadsheet integrity; this aspect is more fully reported in industry based studies (see KPMG survey reported above). However, more importantly from the teaching view, Panko goes on to state:

*"These studies… used a variety of subjects from rank novices to highly experienced spreadsheet developers. All subject groups made errors, and when Panko and Sprague [1999]\* directly compared error rates for undergraduates, MBA students with little or no spreadsheet development experience, and MBA students with at least 250 hours of spreadsheet development experience, they found no significant differences in error rates across the groups"*

*\*Reference to* Panko, R. R., and Sprague, R. H. J. "Hitting the Wall: Errors in Developing and Code-Inspecting a 'Simple' Spreadsheet Model." *Decision Support Systems*, (22) 1999, pp. 337-353.

## 2.0 THE CAUSES OF ERRORS DURING TRAINING

There is a common problem with students learning new computing skills. Research conducted by Van Vliet et al [2] into comparing self-appraisal and objective tests of learners abilities has indicated that novices of both sexes consistently overrate their own computer literacy skills. This has led them to think they have correctly solved a problem when, in fact, they have not.

Panko R.[12], cites one experiment in which student spreadsheet developers were given a spreadsheet to build from a written specification. The 'developers' were then asked to estimate the likelihood that they had made an error during development. The median estimate was 10%, and the mean was 18%. In fact, 86% had made an error in their spreadsheet. When debriefed in class and asked to raise their hands if they thought they were among the successful 14%, well over half of all subjects raised their hands.

Research conducted by Chadwick et al. [3] has also shown that student spreadsheet builders frequently make errors that they are quite confident are not there and are often amazed when the errors are pointed out to them. The same research has catalogued the different types of common errors and has shown that, amongst student spreadsheet builders, there is a tendency to create formulae that do not accurately model the calculations and processing phenomena of the real-world, and to copy formulae incorrectly from one part of a spreadsheet model to another. An assertion made in Chadwick et al. [3] is that even when training is given it too frequently concentrates on 'how to do things correctly' and often ignores 'how to avoid doing things incorrectly'. One possible way of 'avoiding doing things incorrectly' is to make novice spreadsheet builders more aware of common errors and encourage them to apply checking controls during development commensurate with the amount of risk associated with producing a possibly incorrect model. This is supported in the professional audit literature reviewing the teaching of computer auditing:

*"In terms of teaching …several activities can help students …understand that control should be used sparingly but appropriately; the right amount of control depends on the associated risk"* Herremans [4]

During the exercise reported in this paper, students were specifically asked to assess the risks inherent in any model they built. Error-awareness skills were also extensively developed by providing opportunities for students to check for errors :

1. through self-assessment by error-auditing their own spreadsheet models and also those deliberately seeded with errors for training purposes,

2. through peer-assessment by auditing the models of other students.

## 3.0 SELF AUDIT

Self-assessment performed by students has been gaining in popularity in the last few years as a means of enabling students to be reflective on the quality of their own work. Such an approach has been found to be effective in improving intellectual internalisation and intuition [9]. During the trials reported herein self-assessment techniques were based upon the self-audit methods in use in industry. Formal lecture sessions were given in which students were encouraged to consider the advantages and disadvantages of spreadsheet audit in general. The main lecture themes were designed to improve awareness of:

- the usefulness of auditing spreadsheet models,

- the need to find errors, their causes and effects,
- the business risks associated with incorrect spreadsheets,
- the common errors by being exposed to them through seeded models,
- the advantages of using self-audit check lists during development.

### 3.1 Improving Awareness of the Usefulness of Self-Auditing

Students were made aware that Control Self Assessment (CSA) techniques were accepted audit methods in industry. They were asked to consider the advantages and disadvantages of performing self-checking as in fig 3.1.

Fig 3.1 Control Self Assessment (CSA)

| CSA: Advantages | CSA Disadvantages |
| --- | --- |
| Can build CSA into own work schedule | May be tendency not to do at all |
| Learn by detecting own errors | May be that not all errors get detected |
| Greater awareness so fewer errors later | May become blasé and so negligent |
| Etc … | |

### 3.2 Improving Awareness of Errors, Their Causes and Effects

Students were encouraged to identify the events during development which could create, propagate or exacerbate errors. They were asked to tabulate what they thought were possible error-events, their causes and possible effects as in fig. 3.2.

**Fig. 3.2 Sample Error-Event Table : Error Events, Causes, Effects**

| Error -inducing events identified? | Causes? | Effect ? |
| --- | --- | --- |
| Incorrect Formulae | 1. Typographical error <br> 2. Wrong arithmetical precedent rules | Wrong output data leading to faulty decision-making |
| Mistakes in input data-set | 1. Typographical error <br> 2. Wrong data-set <br> 3. Wrong data source | Wrong output data leading to faulty decision-making |

### 3.3 Improving Awareness of Business Risks

As part of their introduction to CSA, students were told that organisations often extended CSA to consider the inherent risks (Control and Risk Self-Assessment or CRSA). CRSA, in this context, involved personal awareness of the risks inherent in a spreadsheet project, the possible error situations, alternative arrangements a client may need in event of failure as well as the developer's own response to failure and problems. Students were encouraged to identify the possible risks inherent in an application they had been asked to develop and to identify the minimum risk-set.

Fig 3.3 Minimum Risk Set Compiled By A Typical Student

| Risk Self Assessment of Errors In My Spreadsheet |
| --- |
| How important to my client is the spreadsheet I am developing ? |
| Which parts of the spreadsheet model are most critical to my clients business? |
| What compensation might the client demand from me for any error ? |

### 3.4 Improving Awareness Using Specially Seeded Models For Training

Improving awareness of errors was accomplished by presenting students with spreadsheet models deliberately created with errors and asking them to identify the errors and assess the extent of commercial risk associated with them. An introductory example from the teaching materials is shown in 3.4.1 (there were more complicated models!).

### 3.4.1 Example of An Error Seeded Model

*Bungee Breakspeare wants to open his own night-club and has put details of expected income and out-goings in a spreadsheet which he has sent to his bank-manager for a start-up loan.(see fig3.4).*

*Question: Which parts of the spreadsheet carry the most likelihood of an error ?*

*Question: Is there an error in this spreadsheet ?*

*Question: What business risks are associated with an error in this model?*

Fig. 3.4 Example of Error identification and Risk Assessment question

| Figures in £ | Jan | Feb | Mar | Apr | May | June |
|---|---|---|---|---|---|---|
| Entrance Ticket sales | 4,000 | 3,500 | 3,000 | 3,000 | 4,000 | 5,000 |
| Bar sales: drink and food | 1,500 | 1,000 | 1,000 | 1,000 | 2,000 | 2,000 |
| **TOTAL INCOME** | **5,500** | **4,500** | **4,000** | **4,000** | **6,000** | **7,000** |
| Wages: bouncers and bar staff | 1000 | 1,000 | 1,200 | 1,200 | 1,500 | 1,500 |
| Electricity charges | 0 | 3,500 | 0 | 0 | 4,000 | £0 |
| Rent for premises | 0 | 0 | 5,500 | 0 | 0 | 5,500 |
| **TOTAL OUTGOINGS** | **1000** | **4500** | **6700** | **1,200** | **5,500** | **7,000** |
| **MONTHLY PROFIT** | **4,500** | **0** | **-2,700** | **2,800** | **500** | **0** |
| **ACCUMULATING PROFIT** | **4,500** | **4,500** | **-2,700** | **100** | **600** | **600** |

### 3.5 Practical Self Assessment Check List

The use of check lists is well documented in audit literature. Their use has also been championed in educational studies with business students learning to build spreadsheet models [8]. The checklist used with students in the research reported herein was a pre-event checklist asking students to consider appropriate action prior to building a spreadsheet.

Fig. 3.5 The Self-Assessment checklist stages

> Before starting your spreadsheet have you considered if and when to perform :
> 1. the RADAR development stages yes/no?
> 2. modularisation of the spreadsheet if necessary yes/no ?
> 3. creation of a logical model of the spreadsheet yes/no ?
> 4. use of the 2 A's : which functions to test yes/no ?
> 5. create a User-Guide yes/no ?
> 6. create a risk assessment table identifying risks yes/no ?

(RADAR life-cycle, spreadsheet modularisation, logical models, and 2-A's approach are spreadsheet building methods developed and taught at the University of Greenwich - for further information on some of these terms refer to Chadwick et al [5] and Rajalingham et al. [6] and [7]. For risk assessment see 3.3 above)

### 4.0 PEER ASSESSMENT

Peer assessment is frequently used as an educational tool because "of its potential for the development of students' autonomy, maturity and critical abilities" [10] . The advantages are cited as giving students appreciation of other students' work and thereby identifying common mistakes. However, there are generally two main criticisms from the students' point of view. It is commonly stated that although the peer assessor is given the opportunity to review another students work, they, the peer assessor do not :

- gain any particular new skills by so doing,
- have their own proficiency at peer assessment itself assessed,
- understand why they should perform part of the tutor's job.

These factors have been major drawbacks to the use of peer assessment generally. However, as a possible solution to this, the approach herein described uses methods derived from the peer audit functions commonly used in industry. Peer audit is a method promoted by internal audit departments whereby application development projects are checked against corporate standards by persons not involved in the original development but of equal skill and business knowledge to the original developers. Such peer auditors are commonly trained in basic audit techniques and taught how to report their findings. Significant savings have been made by companies employing this approach as it reduces the need for a costly central audit department. It also enables far more frequent reviews to occur and at much earlier stages in the application development life-cycle. A peer assessment method used in education and based on professional peer audit was considered to possibly have certain advantages which would address some of the criticisms given by students above. For the assessing students in particular, it was considered that they might develop a set of skills commensurate with those of practicing auditors and which would be immediately transferable to a work environment. Proficiency in such skills could also be educationally assessed.

**The peer audit skills were taught using lectures and practical exercises revolving around:**

- improving awareness of the usefulness of peer-audit,
- a practical peer audit check list,
- practical guidance on writing a peer-audit report,
- exercises for peer audit practice.

### 4.1 Improving Awareness of the Usefulness of Peer-Audit

Students were introduced to peer-audit and encouraged to tabulate the advantages and disadvantages.

Fig 4.1 Peer Audit Advantages and Disadvantages As Per Typical Student

| PA: Advantages | PA: Disadvantages |
|---|---|
| Peers have similar spreadsheet skills | But may make same mistakes |
| Peers have similar domain knowledge | But it may be too limited |
| Etc… | |

### 4.2 Practical Peer Audit Check List and Peer Audit Report

Three types of practical audit methods were used to develop skills in searching for and identifying errors and omissions in other students' work. Peer assessing students were encouraged to:

- find errors in other students spreadsheets using a post-event check list,
- prepare an audit report on their findings.

The post-event check list (fig. 4.2) was used by the assessing student to check that the assessed student had actually done what was required in terms of the standards of model development defined in the requirements specification given to them.

Fig. 4.2 Peer Audit Check List and Report Contents

> **Peer Audit Check List**
> 1. Check User-guide in existence with minimum information
> 2. Check spreadsheet has been modularised correctly
> 3. Select and check key functions using 2 A's approach
> 4. Test several key functions with sample input data from the user
>
> **Peer Audit Report**
> Prepare audit report (one side A4) includes auditors name, date, identifying details of spreadsheet, auditors findings on above 4 checks.
> The audit report is to be a WP document with same name as spreadsheet file.

## 4.3 Exercises In Use of the Peer Audit Method

The peer audit method was used for practical assessment of each student's spreadsheet. The peer exercise occurred in a university computing laboratory with a working version of the spreadsheet. This was in accordance with the work of Boud (in Heywood, 1989) who argued;

*"peer assessment should be formalised in the laboratory. In this way students begin to take responsibility for their own learning and gain insight into their own performance through having to judge the work of others.* **[11]**

Fig. 4.3 Peer Audit Sheet giving instructions on how to perform audit

> **AUDITOR STUDENT NAME:……………**
> **AUDITED STUDENT NAME:……………**
>
> **PART A : AUDIT OF SPREADSHEET 1**
> 1 mark if correct, 0 if spreadsheet is wrong
>
> **Qu 1 :** Change the $ to £ exchange rate to 1.69, Change the D.Mks to £ exchange rate to 2.40, Change the Yen to £ exchange rate to 1300
> The 3 Mthly Total (in I9) should change to 4773.99
>
> **PART B : AUDIT OF SPREADSHEET 2**
> 1 mark if correct, 0 marks if wrong or non-existent
>
> **Qu 2 :** What is the figure for the 3 Mthly Total over all salespeople for April to June? Should be 47425.00
> **Qu 3** Does User Guide identify spreadsheet builder?
> **Qu 4** Does User Guide show the Date of Creation?
> **Qu 5** Does User Guide specify spreadsheet purpose?
> **Qu 6** Is there a lookup table for commission rates?
> **Qu 7** Check any 3 key functions. Are the functions correct?
> **MARK GIVEN:**
>
> Now Prepare Your Audit Report on Spreadsheet 2 model.

Each student was given a written requirements specification and given time to build the appropriate spreadsheet model. The model was then audited by another student using the peer audit sheet (sample shown in fig. 4.3) who gave a mark and wrote an audit report upon his/her findings. The audited student received the mark and the auditing student was themselves graded by the tutor on the quality of their audit report.

To avoid collusion and bias, it was deemed necessary to select the auditing 'pairs' by producing a list of names, randomly sorted, and requiring each to audit the name beneath theirs in the list with a wrap from bottom to top of the list. This ensured that:-

- students did not audit their personal friends,
- no mutual-audit pairs arose i.e. A audits B, B audits A,

- no mutual-audit triples arose i.e. A audits B, B audits C, C audits A.

## 5.0 CONCLUSIONS

The audit approach was used on a cohort of second year undergraduate information systems students. These fairly novice students (in serious spreadsheet work) were selected for this project for two reasons. They were generally eager to participate in new assessment methods and also had not yet formed bad habits which resulted in poor spreadsheet model design. This is in keeping with the suggestion by Habeshaw et al. [10] and also by Chadwick et al. [1] that peer assessment techniques should be started early in the programme of study.

### 5.1 Analysis of Student Feedback On The Exercise

Anonymous feedback was obtained from 42 participating students who answered five questions and were permitted to give comments. All reported favourably on most criteria (except report writing) saying they had not only learned from the experience but also enjoyed it.

Fig.4.4 Feedback from students

|  | Yes | No/Not sure |
|---|---|---|
| 1.SELF AUDIT: was the exercise useful in exposing your mistakes? | 50% | 50% |
| 2.SEEDED MODEL: were these useful in increasing your ability to find mistakes? | 65% | 35% |
| 3.PEER AUDIT: was the exercise useful in developing your skills for finding errors in other people's work? | 72% | 28% |
| 4.PEER REPORT: was the exercise useful in developing your report writing skills? | 45% | 55% |
| 5.PEER AUDIT: was being audited a learning experience for you? | 55% | 45% |

*SELF AUDIT: was the exercise useful in exposing your mistakes?*

Students commented favourably on the disciplined approach forced upon them. Although they identified the time overhead of having to think about inherent risks they felt that the insight gained into possible problem areas outweighed the time overhead.

*SEEDED MODEL: were these useful in increasing your ability to find mistakes?*

Students were keen on this method of raising error-awareness and requested more exercises of further complexity.

*PEER AUDIT: was the exercise useful in developing your skills for finding errors in other people's work?*

Student were enthusiastic that having to audit each other's work had been a beneficial exercise. Many remarked how surprised they were to see the variety of spreadsheet models produced from what they had considered to be requirements that only had 'one' solution.

*PEER REPORT: was the exercise useful in developing your report writing skills?*

The seemingly negative reply to this question may indicate that the question was wrongly asked on the follow-up questionnaire. It may well have been that the report was not useful in 'developing' skills already in existence (all the students had already received much exposure to report writing opportunities).

*PEER AUDIT: was being audited a learning experience for you?*

There is no doubt, from the feedback received, that students were motivated to do well by the prospect of having their work examined by another.

### 5.2 Analysis of The Complete exercise

Crucial to the success of this exercise were:

- early explanation of the rationale for the peer and self-assessment,
- clear and simple instructions for the students to follow,
- objective criteria using prescribed test data, and
- monitoring of the exercise by the tutor through viewing the audit report.

The work continues with current cohorts of students and the methods used have been refined. The concept of peer and self audit has proved to be so useful with spreadsheet teaching that it is now being extended to the teaching of databases.